\documentclass[aps,prl,twocolumn,groupedaddress,showpacs]{revtex4}
\usepackage{graphicx}
\usepackage{isolatin1}
\newcommand{\myscalebox}[1]{\scalebox{0.42}[0.42]{#1}}
\newcommand{\be}{\begin{equation}}
\newcommand{\ee}{\end{equation}}

\begin{document}
\title{ Partition function of two- and three-dimensional Potts
  ferromagnets for arbitrary values of $q>0$.}
\author{A. K. Hartmann}
\affiliation{Institut f\"ur Theoretische Physik, Universit\"at G\"ottingen,
Friedrich-Hund-Platz 1, 37077 G\"{o}ttingen, Germany}
\date{\today}
\begin{abstract}
A new algorithm is presented, which allows to calculate 
numerically the partition function $Z_q$ of the $d$-dimensional
$q$-state Potts models for arbitrary real values $q>0$ at any given temperature
$T$ with high precision. The basic idea is
to  measure the distribution of the number of
connected components 
in the corresponding Fortuin-Kasteleyn representation and to
compare with the distribution of the case $q=1$ (graph percolation),
where the exact result $Z_1=1$ is known. 
As application, $d=2$ and $d=3$-dimensional ferromagnetic Potts models are
studied, and the critical values $q_c$, where the transition changes
from second to first order, are determined. 
Large systems of sizes
$N=1000^2$ respectively $N=100^3$ are treated. The critical value
$q_c(d=2)=4$ is confirmed and $q_c(d=3)=2.35(5)$ is found.
\end{abstract}
\pacs{05.50.+q,05.10.-a,75.10.-b}
\maketitle

The partition function is a quantity of fundamental importance,
because it describes completely the behavior of any
statistical physics model. Unfortunately, in finite-dimensions, only few
models are analytically tractable \cite{baxter1982}. Hence, 
Monte Carlo (MC) simulations \cite{newman1999,landau2000} are usually applied. 
The standard approach to
obtain the partition function is to measure the free energy 
by thermodynamic integration of the
specific heat, i.e. the fluctuations of the energy.
Since this approach is based on measuring fluctuations, 
it is not very efficient, hence limited to
small sizes. One can speedup simulations for certain types of systems 
by applying cluster
algorithms \cite{swendsen1987,wolff1989,chayes1998}, multi-histogram
methods \cite{ferrenberg1989}, multicanonical simulations
\cite{berg1992,janke1995} or transition-matrix
Monte Carlo \cite{wang2002}, 
but the general problem of the strong fluctuations remains. 
To overcome this problem, 
recently Wang and Landau introduced \cite{wang2001} a 
simple yet very efficient method to obtain the partition function.
The key idea is to measure the density of states by performing a biased random
walk in energy space via spin flips.
 It works well for unfrustrated  systems,
e.g. the standard $q$-state Potts model \cite{potts1952}, 
which has become a
standard testing ground for Monte Carlo algorithms. The Potts model
is of profound interest,
because, for dimensions $d$ larger than one, it exhibits order-disorder
phase transitions \cite{wu1982}, which are of second order for $q$
smaller than a critical value $q_c(d)$, while they are of first order for
$q>q_c(d)$. It is analytically proven \cite{baxter1973}
that $q_c(2)=4$, but e.g. for $d=3$,
the exact value of $q_c$ is not known. From various analytical work
\cite{nienhuis1981,kogut1982,grollau2001} and simulations of moderate-size
systems \cite{lee1991,barkema1991,gliozzi2002}, $2<q_c(3)<3$ seems likely.
Unfortunately, since the Wang-Landau method is based on spin flips, 
it works only for integer values of $q$, hence the partition function
for  $2<q<3$ cannot be obtained for large systems in this way.

In this letter, an algorithm is presented, which allows to calculate 
numerically the partition function $Z_q$ of the $d$-dimensional
$q$-state Potts models for arbitrary real values $q>0$ at any given temperature
$T$ with high precision for large systems. The basic idea is
to  measure the distribution of the number of
connected components in the corresponding Fortuin-Kasteleyn (FK) representation
\cite{fortuin1972} and to
compare with the distribution of the case $q=1$ (graph percolation),
where the exact result $Z_1=1$ is known. Large system like $N=1000^2$
resp.\ $N=100^3$ can be treated here, because
for the MC simulation, the cluster algorithm of Chayes and Machta
\cite{chayes1998} is applied. Using this combined approach the critical value
$q_c(d=2)=4$ is confirmed and $q_c(d=3)=2.35(5)$ is found. The outline
of the paper is as follows. Next, the model is defined. Then the
algorithm for calculating the partition function is presented. In the
main part, the results for the $d=2$ and $d=3$ Potts models are
shown. Finally a summary is given.

The $q$-state Potts model \cite{potts1952} 
for integer values of $q$ consists of $N$ spins
$\sigma_i\in \{1,\ldots,q\}$ living on the sites of an arbitrary graph
or lattice $G$, with the Hamiltonian
$H = - \sum_{(i,j)} \delta_{\sigma_i,\sigma_j}$,
where the sum runs over the edges $(i,j)$ of $G$, 
and $\delta$ is the Kronecker delta. For $G$, here $d$-dimensional
hypercubic lattices having periodic boundary conditions 
with nearest-neighbor interactions are considered. The partition function
$Z_q=\sum_{\{\sigma_i \}}e^{-H/T}$, $T$ being the 
temperature, can be written in the FK representation
\cite{fortuin1972} as
\begin{eqnarray}
\label{eq:Z}
Z_q & = &\sum_{G^\prime \subset G} W_q(G^\prime) \nonumber \\
& \equiv &\sum_{G^\prime \subset
  G} p^{N_b(G^\prime)}(1-p)^{N_b(G)-N_b(G^\prime)}q^{N_c(G^\prime)}\,,
\end{eqnarray}
where the sum runs over all subgraphs of $G$ having the same set of
sites and any subset of edges, $W_q(G^\prime)$ is the weight of graph
$G^\prime$, $p=1-e^{-1/T}$,
$N_b(G)$ resp. $N_b(G^\prime)$ are the number of edges in $G$ resp.\ $G^\prime$
and $N_c(G^\prime)$ is the number of connected components in $G^\prime$. The
FK representation allows for  an extension of the model to arbitrary
real values of $q>0$.

For $q=1$, $W_1(G^\prime)$ is the probability of the subgraph
$G^\prime$, if the graph is generated randomly by making
every edge a member of the subgraph with probability $p$. 
This allows for a very efficient generation
of graphs distributed according to $W_1$, i.e. by importance sampling.
Since one generates each time with probability one {\em
some} random subgraph in this way, one has trivially $Z_1=1$.

This allows for a calculation of the partition function $Z_q$ for any
$q>0$ in the following way. Let $P_q(c)$ the probability to have $c$
connected components in a subgraph generated according to weight $W_q$. Then we
have by definition

\begin{eqnarray}
P_q(c) & \equiv &  \frac{1}{Z_q} \sum_{G^\prime \subset G} 
              W_q(G^\prime) \delta_{N_c(G^\prime),c} \nonumber \\
& = &  \frac{1}{Z_q} \sum_{G^\prime \subset G} 
          W_1(G^\prime)q^{N_c(G^\prime)}\delta_{N_c(G^\prime),c} \nonumber \\
& = &  \frac{q^c}{Z_q} \sum_{G^\prime \subset G} 
          W_1(G^\prime)\delta_{N_c(G^\prime),c} \nonumber \\
& = &     \frac{q^c}{Z_q} Z_1P_1(c) = \frac{q^c}{Z_q} P_1(c)\,.
\end{eqnarray}
Hence, we get
\be
\label{eq:Z_Pc}
Z_q = q^c\frac{P_1(c)}{P_q(c)}\,.
\ee
This means, by measuring the probability distributions of the number
of connected components for random subgraphs ($q=1$) 
and for the target value $q$, we can obtain
$Z_q$. Note that Eq.\ (\ref{eq:Z_Pc}) holds for {\em all} values of
$c$ simultaneously \cite{remark1}. Therefore, by comparison of the full
distributions, one has a mean to determine $Z_q$ with very high precision
.

Eq.\ (\ref{eq:Z_Pc}) might be useful for analytical calculations, but for
most interesting graphs $G$, the distributions cannot be obtained in this way. 
Hence, one uses
numerical simulations to obtain the distributions $P_1(c)$ and $P_q(c)$.
In practice, one can generate random subgraphs according to $W_1$ with
importance sampling, as explained above, 
and according to $W_q$ using the very efficient cluster algorithm
of Chayes and Machta \cite{chayes1998}. This algorithm allows for
simulation for arbitrary values of $q$, similar to other approaches
\cite{sweeny1992,weigel2002,gliozzi2002}.

Nevertheless, 
for large values of $q$ and finite statistics, $P_1(c)$ and $P_q(c)$ 
will not overlap, because deviations from the typical value are
exponentially suppressed. In this
case one has to study intermediate values $q_1,\ldots,q_k \in [1,q]$,
calculate each time $P_{q_i}(c)$ and $Z_{q_i}$. This allows to extend
$P_1(c)$ stepwise \cite{align,align_book} 
for larger values of $c$, until $P_1(c)$
and $P_q(c)$ have sufficient overlap. In principle, it is a bit ugly,
that one has to perform simulations at several values of $q_i$, but on
the other hand, one gets the partition function for all considered
values, which will be useful in the following. 
Note that for the
Wang-Landau algorithm, also one long run is sufficient only in theory,
in practice, if the system size is larger than tiny, one has to divide
the energy range into intervals, 
perform independent runs for each interval, and match the
results of the different runs as well. Anyway, this is no problem for
either method, because it can be done automatically 
by a program, no matter how many intervals have to be matched.
The real advantage of the present approach
is that it works for all values of $q>0$, since it does
not rely on flips of spins.

\begin{figure}[htb]
\begin{center}
\myscalebox{\includegraphics{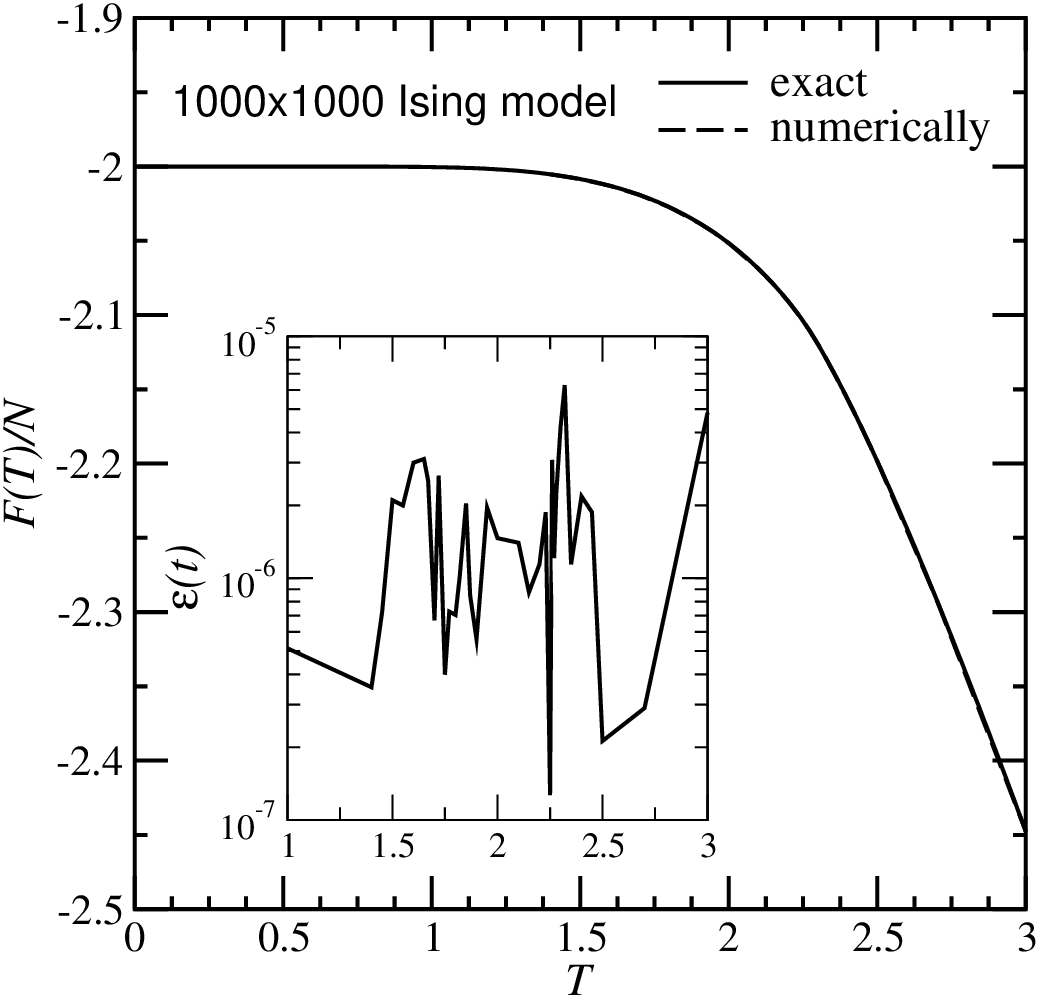}}
\end{center}
\caption{Free energy $F$ per spin as a function of the temperature $T$
of the two-dimensional Ising model
  ($q=2$) obtained by the algorithm and by an exact calculation for
  system size $L=1000$. The inset shows the relative error
  $\epsilon(T)$. For each independent value of temperature, only a total of 
$5.5\times 10^5$ MC sweeps were performed. }
\label{figF2d}
\end{figure}

To test the new approach, it is now applied to 
the two-dimensional Ising model ($q=2$), 
where exact results are available for finite
system sizes \cite{ferdinand1969}.
In Fig.\ \ref{figF2d}, the Gibbs free energy per spin 
$F/N\equiv -\frac{T}{N} \ln Z_q$
is shown in the Ising representation (i.e.\ for the Hamiltonian
$H=-\sum_{(i,j)} [2\delta_{\sigma_i,\sigma_j}-1]$). The data
of the simulation and the analytical result are given for a large system size
$N=1000\times 1000$. Thus, $k=110$ different values $q_i$ are
necessary for measuring $P(c)$ over the desired range. Equilibration
of the cluster MC simulation is determined by monitoring 
the number of connected
components and the number of edges when starting with a full
resp.\ empty subgraph. Equilibration is assumed, when the values for
the different starting conditions agree within the range of
fluctuations. Due to the global update nature of the Chayes-Machta algorithm,
this is the case for typically few Monte Carlo
sweeps. Hence, for each value of $q_i$, $5\times 10^3$ steps where
sufficient, to obtain a high accuracy, as shown in the inset of
Fig.\ \ref{figF2d}.

\begin{figure}[htb]
\begin{center}
\myscalebox{\includegraphics{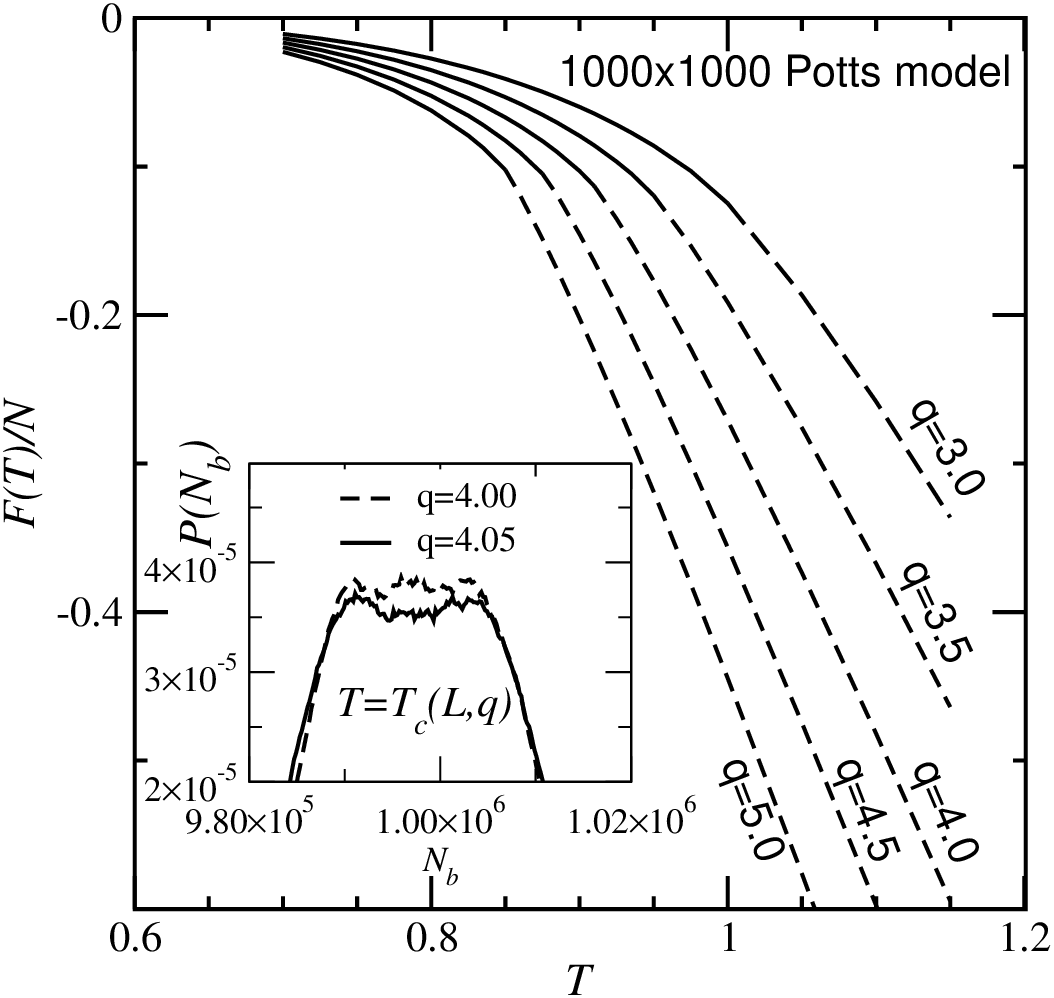}}
\end{center}
\caption{Free energy per spin of the two-dimensional Potts model
  ($3\le q\le 5$) for system size $L=1000$. The inset shows the
  distribution of the number of edges at the  
  temperatures $T(q=4.00)=0.910289$ resp. $T(q=4.05)=0.906865$.  }
\label{figF2dPotts}
\end{figure}

Since the aim is to determine $q_c$, the new approach is further
tested by applying it to the
two-dimensional Potts model, where $q_c=4$ is known.
In Fig.\ \ref{figF2dPotts}, the Gibbs free energy per spin 
is shown for values in the range $3\le q
\le 5$. Due to the large system size, in total $k=261$ different 
values for $q\in [1,5]$ are necessary. 
For $q>4$ a kink at the transition temperatures is visible, as
expected. One could in principle take derivatives of the free energy to
obtain e.g. average energy and specific heat, e.g. to calculate
critical exponents in the case of second order transitions. A better way is to
calculate these derivatives analytically from Eq.\ (\ref{eq:Z}), which
allows to express the mean energy resp.  the
specific heat by the average number of edges resp. 
the fluctuations of the number of edges
\cite{fortuin1972,janke1995}, 
which are available directly from the simulation. 
Since this is a standard approach, it is not further pursued here.

One can determine $q_c$ more precisely by considering the
distribution of the number of edges \cite{lee1990,lee1991,gliozzi2002}, 
see inset of
Fig.\ \ref{figF2dPotts}.  The distributions are obtained by performing several
long simulations for $T \in [0.906,0.907]$
resp. $T \in [0.9100,0.9105]$, exhibiting a total of more than $2\times 10^7$ MC
sweeps for each value of $q$, 
and combining the results from different temperatures
using the multi-histogram approach \cite{ferrenberg1989}, see also
Ref. \cite{newman1999}.
For $q=4.05$ a
two-peak structure is visible, while for $q=4.00$ not. This confirms
within the given numerical accuracy the known result $q_c=4$.

\begin{figure}[htb]
\begin{center}
\myscalebox{\includegraphics{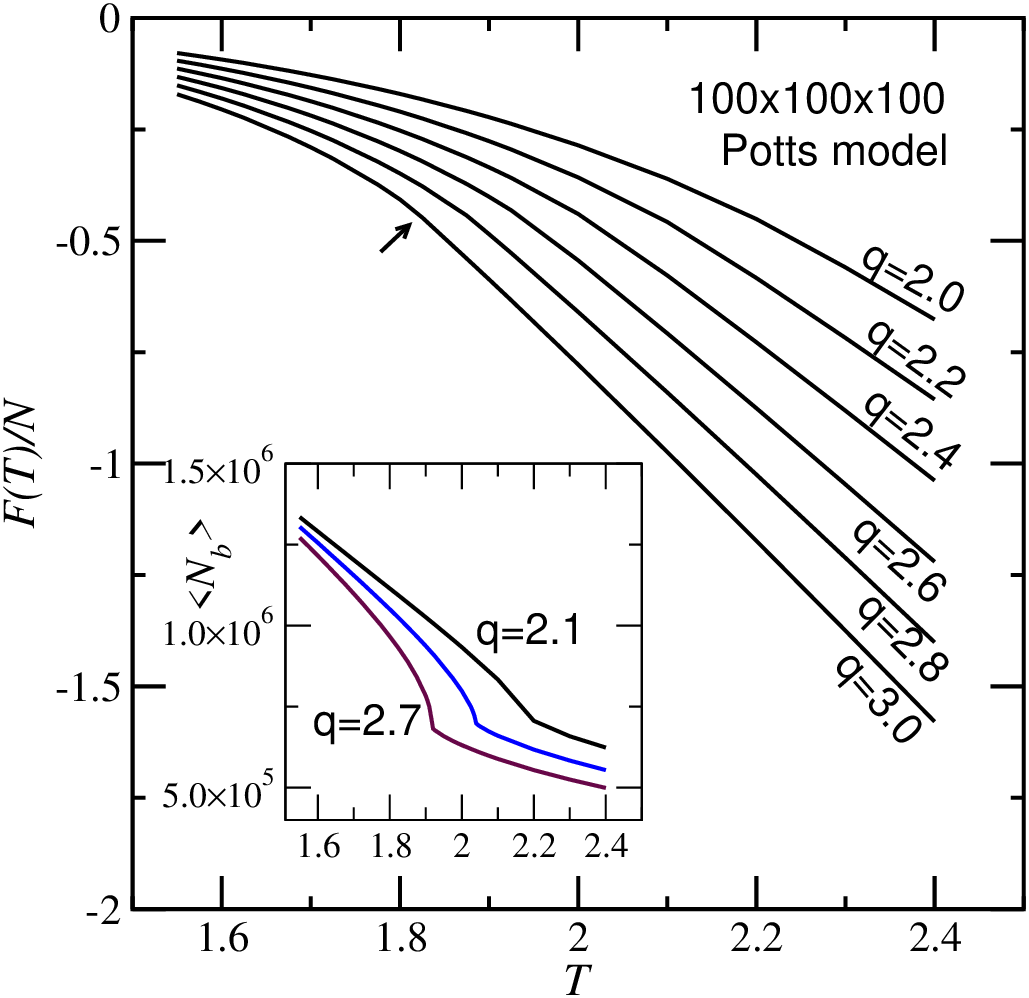}}
\end{center}
\caption{Free energy $F$ per spin of the three-dimensional Potts model
  ($2\le q\le 3$) as a function of the temperature $T$
for system size $L=100$. The arrow indicates the
  phase transition for $q=3$. The inset shows the average number of
  edges $N_b$ as a function of $T$ for $q=2.1, 2.4$ and $2.7$.}
\label{figF3dPotts}
\end{figure}

\begin{figure}[htb]
\begin{center}
\myscalebox{\includegraphics{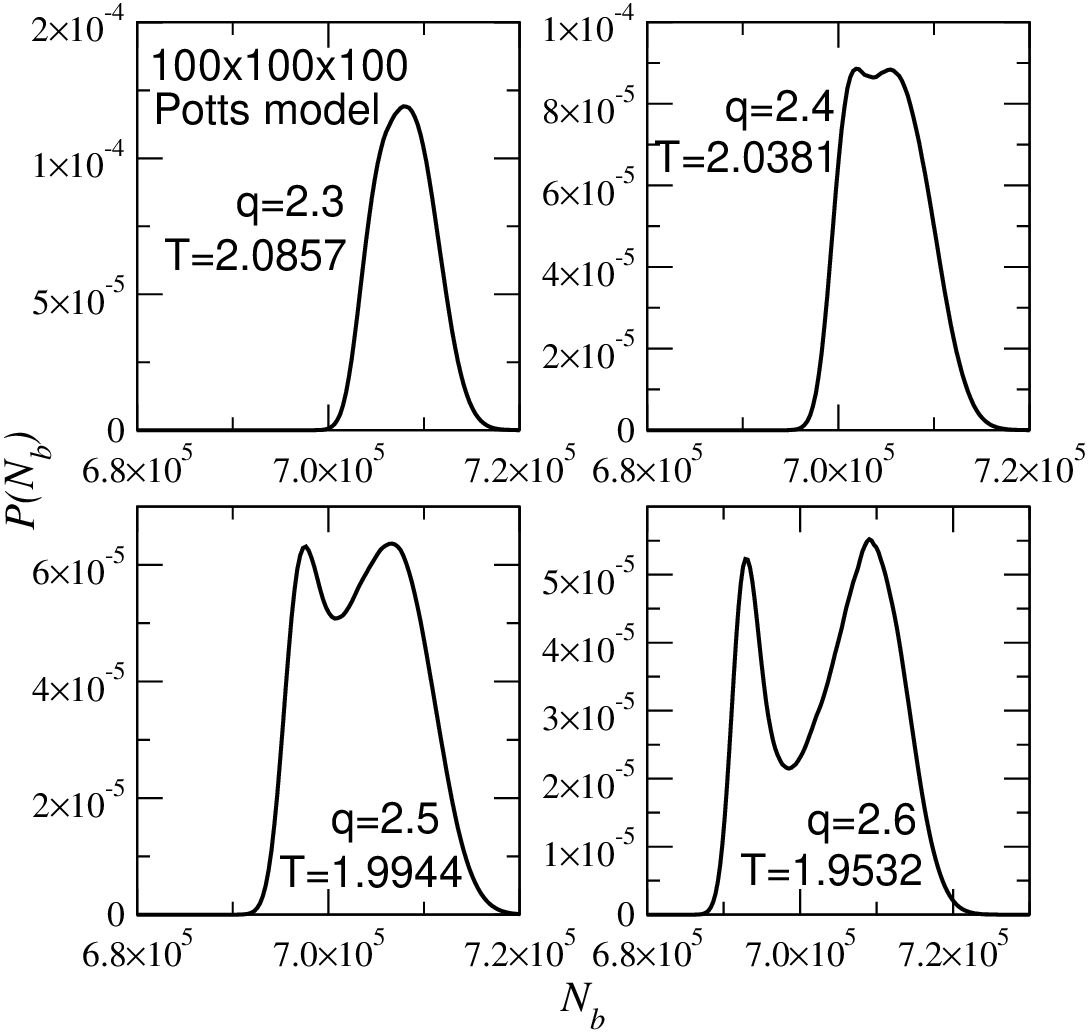}}
\end{center}
\caption{Distributions of the number of edges at temperature 
$T$ close to the critical value $T_c(L,q)$ for
  $q=2.3, 2.4, 2.5$ and $2.6$.}
\label{figDistr3d}
\end{figure}

For three dimensions, the situation is less clear, no exact analytic
results are available. 
A value of  $2<q_c(3)<3$ seems likely, see  analytical work
\cite{nienhuis1981,kogut1982,grollau2001} and simulations of moderate-size
systems \cite{lee1991,barkema1991,gliozzi2002}. In the range where the
transition is first order, the transition seems to be weak,
i.e. making a direct numerical treatment difficult. This is confirmed
by the results for the free energy calculated using the present approach for
$N=100^3$, see Fig.\ \ref{figF3dPotts}.
  The data is obtained by combining the results for $k=212$
different values $q_i\in [1,3]$. 
No clear kink in any of the functions is visible. Thus,
to obtain a precise estimate for $q_c$,
 one has to study again the number of edges. The average, shown
in the inset of Fig.\ \ref{figF3dPotts}, allows to see the transition
point well, but to infer the order of the transition is still difficult
because of finite-size rounding of the curves.
The full distributions close to  the phase transition are presented in
Fig.\ \ref{figDistr3d}. The distributions are obtained again by
performing simulations for several temperatures close to $T_c(L,q)$,
for each value of $q$ more than $2\times 10^6$ MC sweeps, and combining
the data using the multi-histogram approach \cite{ferrenberg1989}.
 For $q=2.6,\,2.5$ one can see a clear double-peak
structure, while for $q=2.4$ the distribution has only a faint
double-peak structure. Clearly just one peak is present for $q=2.3$. 
Since the depth of the minimum between the two peaks grows with system
size \cite{lee1990}, it is still possible that $q_c$ is even below
$q=2.3$, but from the shape of the distribution at $q=2.3$ this seems unlikely.
This allows to conclude $q_c(d=3)=2.35(5)$ from the present results.

This result is smaller than the result $q_c=2.620(5)$ obtained by
Gliozzi. The deviation is probably due to the fact that in that work
much smaller system sizes $N=14^3$ where used, which shifts the value
of $q_c$ up, as discussed above. The result $q_c=2.45(10)$ obtained by
Lee and Kosterlitz \cite{lee1991} is compatible with our result,
although is even  less reliable since small sizes were used and
data obtained at $q=3$ was extrapolated to values $q\in [2.7,3]$. 
Barkema and de Boer studied \cite{barkema1991} 
a model with integer $q$, but mimicking the behavior of any $q>0$, and
got $q_c=2.21$. The
results of analytical studies are scattered around the result obtained
here: Kogut and Sinclair found \cite{kogut1982} $q_c=2.55$ in a $1/q$
expansion, Nienhuis et al. obtained \cite{nienhuis1981} $q_c\sim 2.2$
using a real-space renormalization approach, while Grollau et al. got
\cite{grollau2001} $q_c\sim 2.15$ within a Ornstein-Zernicke Approximation.

To summarize, a new approach to calculate numerically the partition
function of $q$-state Potts models for arbitrary values $q>0$ is
presented. Using a combination with a fast cluster algorithm large
system sizes can be treated. The method is evaluated by performing a
comparison with exact analytic results for two-dimensional Ising
models of size $N=1000^2$, a very good agreement is found. For the
$d=2$ Potts model of the same size, the analytically
obtined critical value 
$q_c=4$ is confirmed. For the three-dimensional
Potts model, due to the weakness of the first order transition, it is
hard to infer $q_c$ from the data (size $N=100^3$) for the free
energy. From the analysis of the distribution of the number of edges
in the generated subgraphs, $q_c=2.35(5)$ is concluded.

The approach to obtain the free energy should be extensible beyond the
standard $q$-state ferromagnetic Potts model, e.g. for random or diluted
ferromagnets, other lattice types and/or higher dimensions. The method should
work in principle also for frustrated systems, but here the efficient 
generation of the subgraphs remains an open problem.

\begin{acknowledgments}
The author thanks E.O. Yewande and W. Janke 
for critically reading the manuscript.
Financial support was obtained from the
{\em VolkswagenStiftung} (Germany) within the program
``Nachwuchsgruppen an Universit\"aten''.
\end{acknowledgments}

\bibliography{partition,remarks_partition}

\begin{thebibliography}{28}
\bibitem[{rem()}]{remark1}
\bibinfo{note}{Note that one could also try to study the probability $P_q(b)$
  to have $b$ edges, but in this case no connection to a trivial case (like
  having $Z_1=1$ here) is possible, because one cannot evaluate $\sum_{G^\prime
  \subset G} q^{N_c(G^\prime)} \delta_{N_b(G^\prime),b}$}.
\end{thebibliography}

\end{document}